\documentclass[aps,pra,twocolumn,a4paper,showpacs,superscriptaddress,floatfix,10pt]{revtex4}

\usepackage{graphicx}
\usepackage{amsmath}
\usepackage{epsfig}
\usepackage{helvet}
\usepackage{amssymb}
\usepackage{epstopdf}
\usepackage{color}
\usepackage{bbold}
\usepackage{cancel}

\begin{document}

\title{A semi-classical limit for the many-body localization transition}


\author{Anushya Chandran}
\affiliation{Perimeter Institute for Theoretical Physics, Waterloo, Ontario N2L 2Y5, Canada}   \email{achandran@perimeterinstitute.ca}
\author{C. R. Laumann}
\affiliation{Perimeter Institute for Theoretical Physics, Waterloo, Ontario N2L 2Y5, Canada}  
\affiliation{Department of Physics, University of Washington, Seattle, WA, 98195, USA}

\date{\today}

\begin{abstract}
We introduce a semi-classical limit for many-body localization in the absence of global symmetries. 
Microscopically, this limit is realized by disordered Floquet circuits composed of Clifford gates.
In $d=1$, the resulting dynamics are always many-body localized with a complete set of strictly local integrals of motion.
In $d\geq 2$, the system realizes both localized and delocalized phases separated by a continuous transition in which ergodic puddles percolate.
We argue that the phases are stable to deformations away from the semi-classical limit and estimate the resulting phase boundary.
The Clifford circuit model is a distinct tractable limit from that of free fermions and suggests bounds on the critical exponents for the generic transition.
\end{abstract}


\maketitle



The central assumption of statistical mechanics is that interactions between particles establish local equilibrium.
It is now believed that this assumption of ergodicity breaks down in isolated quantum systems in the presence of sufficient quenched disorder \cite{Anderson:1958ly,Gornyi:2005lq,Basko:2006aa,Oganesyan:2007aa,Pal:2010gs, Monthus:2010vn,Vosk:2013yg,Huse:2013kq,Serbyn:2013rt,Luca:2013fe,Bauer:2013rz,Imbrie:2014jk,Kjall:2014aa,Vosk:2014eu,Nandkishore:2014rw,Luitz:2014ak,Laumann:2014aa,Nandkishore:2014dq}. 
The fundamental feature of such \emph{many-body localized} (MBL) systems is the slowness of the spread of entanglement \cite{Znidaric:2008aa, Bardarson:2012kl} due to the presence of local integrals of motion (IOM) \cite{Huse:2013kq,Serbyn:2013uq,Nanduri:2014aa,Chandran:2014db,Imbrie:2014jk,Ros:2014nr}.
These local observables remember initial conditions forever and thereby block both equilibration and transport. 
At lower disorder strengths, the system delocalizes and ergodicity is restored.
There is now much theoretical and numerical evidence for this picture in one dimension \cite{Oganesyan:2007aa,Pal:2010gs,Monthus:2010vn,Vosk:2013yg,Huse:2013kq,Serbyn:2013rt,Luca:2013fe,Bauer:2013rz,Imbrie:2014jk,Kjall:2014aa,Vosk:2014eu,Nandkishore:2014rw,Luitz:2014ak,Znidaric:2008aa, Bardarson:2012kl,Serbyn:2013uq,Nanduri:2014aa,Chandran:2014db,Ros:2014nr,Vasseur:2014cs,Agarwal:2014th} and the experimental search in quantum optical and atomic systems is underway \cite{Kondov:2013om,Serbyn:2014ek,Andraschko:2014tg,Vasseur:2014cs}. 

The nature of the two phases and the dynamical transition between them is poorly understood in dimensions greater than one.
In this article, we present a semi-classical theory of the delocalization transition in the absence of global symmetries in arbitrary spatial dimension.
In the semi-classical limit, the transition maps precisely onto classical site percolation.
Each lattice site either blocks the flow of quantum information or permits it to spread.
Clusters of unblocked sites form `ergodic puddles', as such regions locally thermalize.
Blockage sites, on the other hand, host local integrals of motion.
Thus, at large density of blockage sites $p_b$, the system is localized. 
As $p_b$ decreases, the ergodic puddles grow and eventually percolate ($p_b^c$) leading to delocalization.

\begin{figure}[tbp]
\begin{center}
\includegraphics[width=\columnwidth]{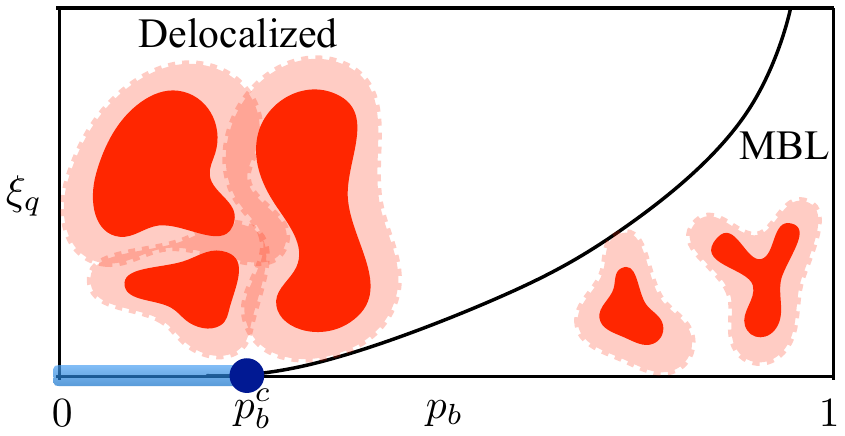}
\caption{Schematic dynamical phase diagram without conservation laws. The semi-classical/Clifford limit corresponds to the horizontal axis ($\xi_q = 0$). The blue line covers the delocalized non-ergodic phase ending at the delocalization transition at $p_b = p_b^c$ in the percolation class. In $d=1$, $p_b^c = 0$. Cartoon: red regions show ergodic puddles with pink regions of influence due to tunnelling.}
\label{Fig:2dphasediagram}
\end{center}
\end{figure}

The sharp distinction between blocked and unblocked sites is a consequence of the lack of quantum tunneling in our semi-classical limit. 
More precisely, the localization length $\xi_q$ of the integrals of motion on blocked sites is identically zero. 
The delocalization transition is instead driven by the divergence of the percolation length $\xi_e$ characterizing the size of ergodic puddles.
Perturbing away from the semi-classical limit, both length scales become important as quantum information may now tunnel through blocked regions.
Fig.~\ref{Fig:2dphasediagram} shows the resulting phase diagram and the generic delocalization transition.

The semi-classical model has several desirable features. 
First, the absence of symmetries distills the underlying entanglement transition from the complicated epiphenomena associated with globally conserved currents.
Second, it is computationally tractable \cite{Nielsen:2000ys} but the phase diagram does not coincide with that of the usual free fermion Anderson models \cite{Evers:2008p10136}. 
Finally, although the phases and transition are semi-classical, they provide a new perturbative starting point for studying the generic transition.

\emph{The model---}
The breakdown of ergodicity is simplest in the absence of global symmetries. 
The equilibrium distribution is unconstrained and there are no slow modes associated with conserved currents. 
We therefore study discrete unitary dynamics generated by the periodic application of a unitary $U$ to a system of $N$ qubits --- that is, the state after $t$ time steps is $|\psi(t)\rangle = U^t |\psi(0)\rangle$.
Such dynamics need not conserve anything, including energy, and are known as Floquet dynamics; see \cite{Lazarides:2014rq,Ponte:2014yo,Abanin:2014ca} for recent studies of MBL in Floquet systems with time-varying Hamiltonians.

\begin{figure}[htbp]
\begin{center}
\includegraphics[width=0.9\columnwidth]{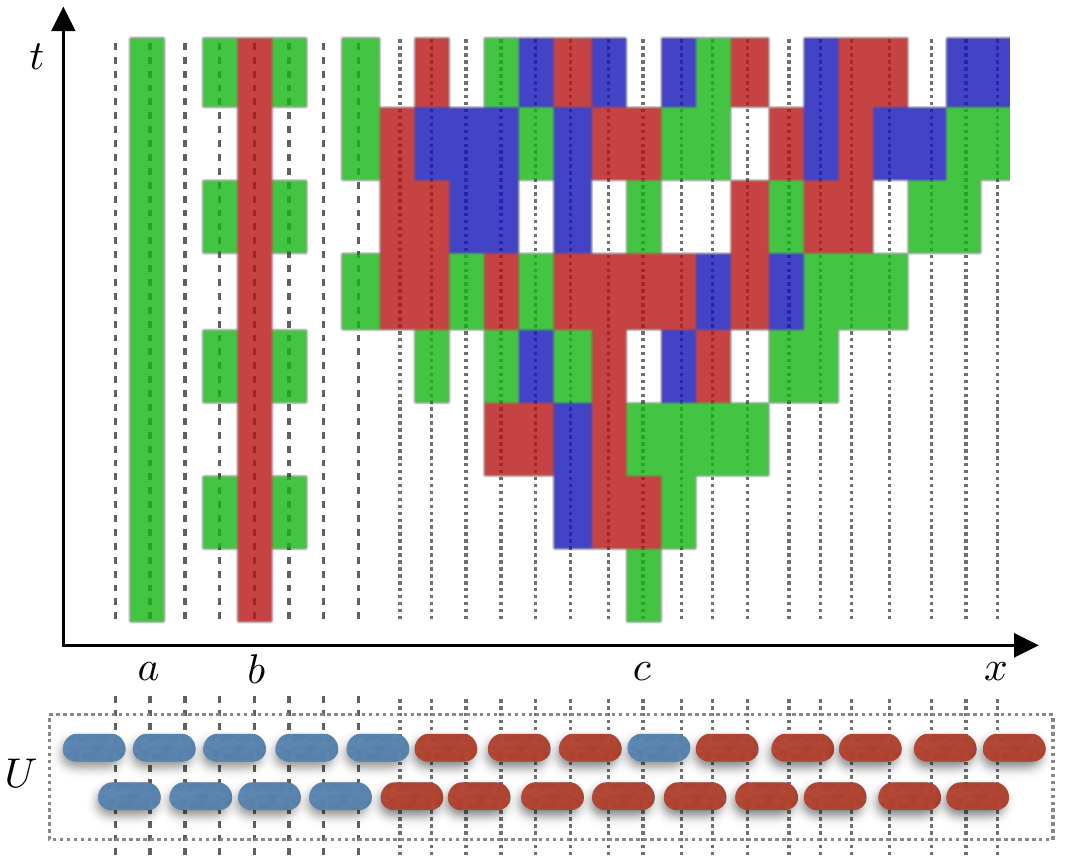}
\caption{
	Space time evolution of Pauli operators $Z_a(t)$, $X_b(t)$ and $Z_c(t)$ under periodic evolution by the $d=1$ brick-layer Clifford circuit $U$ shown below. 
	At each time $t$, the sites are colored red, blue, or green according to the presence of $X$, $Y$ or $Z$ respectively. 
	Dark (light) qubit lines are blockages (non-blockages). 
	Blue (red) ovals denote gate B (R).
	}
\label{Fig:Fig2}
\end{center}
\end{figure}

We implement $U$ as a multi-layer quantum circuit. 
Each layer is composed of non-overlapping nearest neighbor two qubit gates on an underlying spatial lattice, while adjacent layers are stacked in a brick-layer configuration.
An example in $d=1$ is shown at the bottom of Fig.~\ref{Fig:Fig2}. 
In the Heisenberg picture, the support of bounded operators may only grow in discrete steps under the evolution of $U$, which provides a technical advantage over generic time-varying Hamiltonians.

The quenched disorder in the model comes from the choice of the particular two qubit gate at each brick.
Restricting this choice to members of the Clifford group defines our semi-classical limit; we will see below $\xi_q = 0$ for such circuits.
The Clifford group on $N$ qubits consists of those unitaries which preserve the group $\{I,X,Y,Z\}^{\otimes N}$ of Pauli operators under conjugation. 
This is clearly a special subgroup of $U(2^N)$; generic unitaries map Pauli operators into superposition.
Moreover, the Gottesman-Knill theorem \cite{Gottesman:1998aa} guarantees that the time evolution of Pauli operators by such Clifford circuits is classically efficiently simulable.
This provides both analytic and numeric tractability.
For reviews of the many properties of Clifford circuits, see \cite{Gottesman:1998aa,Aaronson:2004db,Nielsen:2000ys}.

Global quenches from product states lead to thermalization in ergodic circuits. 
That is, the reduced density matrix of any finite subsystem $A$ approaches the completely mixed state,
$\lim_{t\rightarrow \infty} \lim_{N\to \infty}\rho_A(t) \propto\mathbb{1}$.
Equivalently, the entanglement entropy of $A$, $S_A(t) = -\textrm{Tr} \rho_A(t) \log_2(\rho_A(t))$ is maximal at late times.
This provides a simple test for ergodicity.
A circuit with a local integral of motion $O$ fails this test as the local density matrices cannot forget the initial value of $O$ even as $t\to \infty$.
We remind the reader that a local integral of motion $O$ is a local operator conserved by the evolution: $[U, O]=0$.

\emph{One dimension---} 
One-dimensional systems are especially prone to localization in the presence of disorder. 
In the Clifford limit, there are local motifs of gates which entirely block the spread of operators in the Heisenberg picture. 
A finite density of such motifs thus chops the chain into puddles from which information about local initial conditions cannot escape, preventing thermalization.

To illustrate this, consider a restricted ensemble:
each gate in $U$ is chosen to be either B or R (blue and red ovals in Fig.~\ref{Fig:Fig2}) independently with probability $p$ and $1-p$. 
The Clifford gates B and R are defined (up to inconsequential signs) by their action on the Pauli generators under conjugation \footnote{The sign of $U O U^\dagger$ for a given Pauli operator $O$ is physically observable but plays no role in thermalization.}:
 gate R: $ ZI \rightarrow ZZ, \,
XI \rightarrow ZY, \,
I Z \rightarrow XX, \,
IX \rightarrow YX$, 
 gate B: $ ZI \rightarrow ZI, \,
XI \rightarrow XZ, \,
I Z \rightarrow IZ, \,
IX \rightarrow ZX
$. 

Figure~\ref{Fig:Fig2} shows three representative Heisenberg evolutions of Pauli operators under the action of a circuit drawn from this gate set. 
There are several distinct behaviors. 
Within the domain of B gates on the left, local operators do not spread. 
The $Z$ operator beginning at site $a$ is conserved by the evolution, representing a strictly local integral of motion. 
The conjugate $X$ operator at site $a$ behaves just like that shown at site $b$: its spreads to the neighboring sites and then oscillates back.
Those sites touched exclusively by B gates are examples of \emph{blockages}.
A blockage site hosts an IOM which remains on-site after each layer within the circuit $U$. 
Elementary arguments show that such sites prevent the evolution of operators with support strictly on the left (right) from extending past to the right (left).

The $Z$ operator beginning at site $c$ spreads uniformly within a light-cone inside the region of R gates on the right. 
This expansion is blocked by the blockage on the left, but passes around the isolated B gate on the right. 
Any other local operator beginning at site $c$ exhibits a similar expansion behavior (not plotted). 
Intuitively, when local Pauli operators evolve into long strings, we expect the reduced density matrix to forget local initial conditions.
Thus, we conjecture that clusters of non-blockage sites act as fully ergodic puddles, wherein $\rho_A(t \to \infty) \propto \mathbb{1}$ for subsystems $A$ less than half the size of the puddle.
We have checked this conjecture numerically for initial product states in which each spin independently points along the $x, y$ or $z$ direction, but we do not have a general proof.
d
We emphasize that blockage sites are not trivial cuts in the chain, as $X$ and $Y$ are not additional integrals of motion.
Rather, the support of $X(t)$ ($Y(t)$) on a blockage site grows until other blockage sites are encountered, as in the oscillations shown by $X(t)$ at site $b$.

Putting the pieces together, the density of blockage sites is $p_b = p^k$ for a $k$-layer circuit drawn from the restricted ensemble.
At any $p_b>0$, these blockages prevent the percolation of quantum information across the chain.
This Clifford localized phase has an extensive algebra of local IOMs of two kinds: the single site IOMs on the blockage sites and larger scale IOMs associated with the ergodic puddles (of typical size $\xi_e$).
Both types of IOMs are \emph{strictly} local ($\xi_q=0$) as they have bounded support on the lattice.
At small $p_b$, the typical puddle size $\xi_e$ diverges as $1/p_b$ and the system is ergodic at $p_b=0$.

\begin{figure}[htbp]
\begin{center}
\includegraphics[width=\columnwidth]{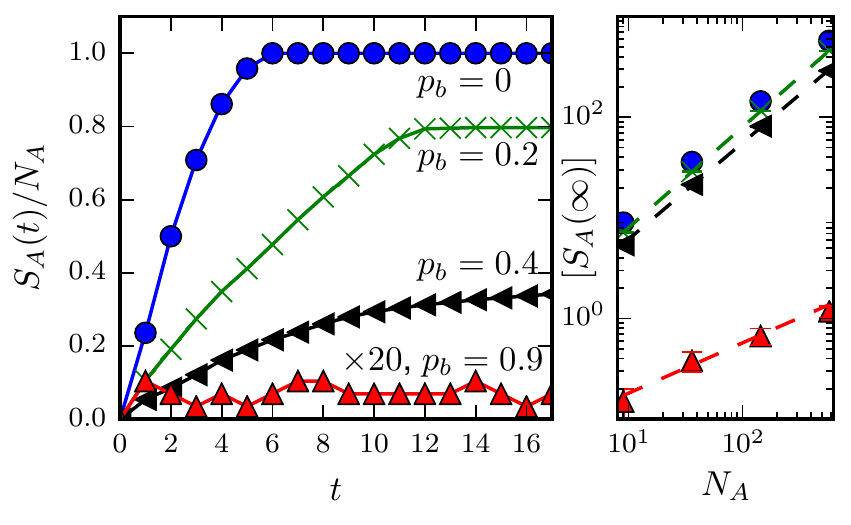}
\caption{(left) Growth of entanglement entropy of subsystem $A$ after global quench from $z$-product state in square lattice Clifford circuit, with $N = 40\times 40$ and $N_A = 24\times 24$.
	(right) Subsystem size $N_A$ dependence of disorder averaged saturated entropy at long times $[S_A(\infty)]$. Dashed lines indicate volume (slope $1$, green), fractal (slope $d_f/d =91/96$, black) and area (slope $1-1/d = 1/2$, red) power laws.}
\label{fig:entgrowth2d}
\end{center}
\end{figure}

\emph{Higher dimensions--- } 
The generalization to $d>1$ is immediate. 
Sites may host blockages, which prevent the spread of operators under Heisenberg evolution. 
If the ergodic puddles of unblocked sites percolate, the circuit delocalizes; if not, it is MBL. 
As usual for site percolation, the percolation threshold $p_b^c$ is non-zero in $d \ge 2$ so that both phases arise in the semi-classical limit with a transition in the percolation universality class.
The associated critical exponents are well known \cite{Shante:1971xi,Isichenko:1992fk}. 
For example, the characteristic size of the ergodic puddles grows as $\xi_e \sim |p_b - p_b^c|^{-\nu}$ as $p_b \rightarrow p_b^c$ from above with $\nu=4/3$ in $d=2$, $\nu\approx0.88$ in $d=3$ and $\nu=1/2$ for $d > 6$, the upper critical dimension.
The MBL phase shrinks in higher dimensions as there are more paths for operators to evade blockage sites.
On the hyper cubic lattice, the metallic clusters stop percolating at $p_b^c \approx 0.4$ in $d=2$, $p_b^c \approx 0.7$ in $d=3$ and $p_b^c \approx 1 - 1/2d$ to leading order in large $d$.

We note that the site percolation problem defined by the blockages is short-range correlated: given that a site is a blockage, its neighbors are slightly more likely to also be blockages.
This detail shifts the percolation thresholds in $d>1$ but does not modify the universal critical properties.

The detailed thermalization properties of the phases in the semi-classical limit are revealed by the growth of entanglement entropy after a global quench (see Fig~\ref{fig:entgrowth2d}).
The strict locality ($\xi_q=0$) of the IOMs in the MBL phase implies that there is no slow growth of entanglement \cite{Znidaric:2008aa, Bardarson:2012kl}, but rather, the entanglement entropy $S_A(t)$ saturates to an area law \cite{Chandran:2014dz} as seen at $p_b = 0.9$. 
In the delocalized phase at $p_b < p_b^c$, the saturated entropy satisfies a volume law ($p_b = 0, 0.2$). 
At $p_b>0$, however, the finite density of blockages suppresses the coefficient relative to the fully ergodic case.
Thus, the semi-classical delocalized phase is not ergodic.

At the critical point, $p_b = p_b^c \approx 0.4$, the intersection of the percolating cluster with subregion $A$ contains $N_p \sim N_A^{d_f/d}$ sites where $d_f$ is the fractal dimension. 
It follows that $S_A(\infty) \sim N_A^{d_f/d}$.
This anomalous growth law at the critical point is rather unusual. 
It is only allowed because the delocalized phase is non-ergodic, as predicted by Ref.~\onlinecite{Grover:2014sy} \footnote{In the semi-classical limit, the growth law of the saturated entanglement entropy after a quench is the same as that of the eigenstates of the circuit $U$. Thus, the results of Ref.~\cite{Grover:2014sy} should apply.}.


\begin{figure}[htbp]
\begin{center}
\includegraphics[width=\columnwidth]{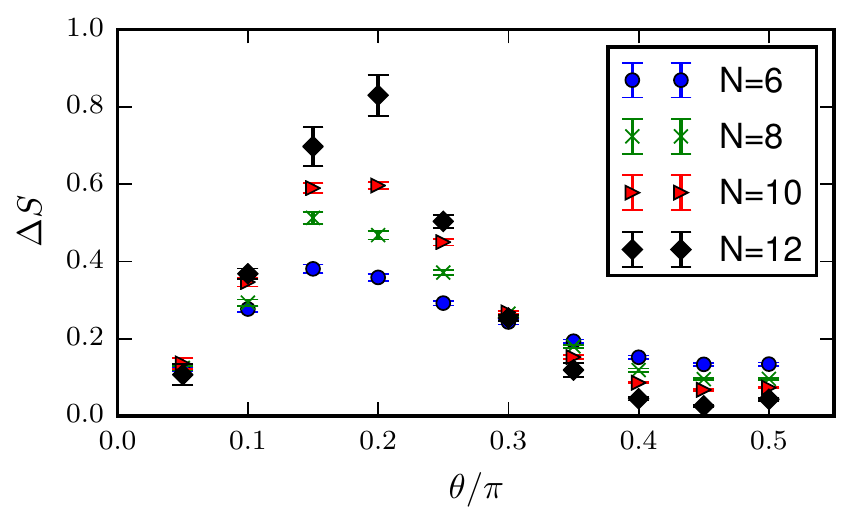}
\caption{Standard deviation of saturated entanglement entropy for perturbed Clifford circuit in $d=1$ at $p_b = 1$. The crossover in finite-size flow at $\theta \approx 0.3 \pi$ indicates the  transition into the delocalized phase at large $\theta$.}
\label{Fig:Fig4}
\end{center}
\end{figure}

\emph{Transition away from the Clifford limit--- }
In this section, we relax the restriction to the Clifford group in order to explore the delocalization transition with $\xi_q \ne 0 $.
This severely limits the analytic and numeric tools available for studying the dynamics: 
the Gottesman-Knill theorem no longer provides a route to efficient classical simulation as even the addition of an infinitesimal single qubit rotation to the set of Clifford gates makes the system universal for quantum computation \cite{Nielsen:2000ys}.
Thus, the arguments to follow are somewhat more heurisic.

We begin by assuming that the Clifford localized phase persists in the presence of weak generic perturbations. 
We will numerically check this assertion for $d=1$ below; there are also several recent works arguing for the existence of Floquet MBL phases in other models \cite{Lazarides:2014rq,Ponte:2014yo,Abanin:2014ca}.
In the perturbed case, the IOMs develop exponential tails with typical length $\xi_q > 0$. 
Near the percolation transition, an IOM associated with an ergodic puddle is of large underlying size $\xi_e$ and the quantum tails simply fuzz the boundary if $\xi_q \ll \xi_e$, see Fig.~\ref{Fig:2dphasediagram}.
The dressed IOMs now interact with one another according to $e^{-d_e/\xi_q}$, where $d_e$ is the distance of closest approach of the puddles.
If the typical interaction scale is smaller than the level spacing on each puddle ($\sim 2^{-N_e}$), where $N_e \sim \xi_e^{d_f}$ is the number of sites in a typical puddle, then the states on the two puddles cannot hybridize and the IOMs remain local.
This suggests the MBL phase persists up to a critical $\xi^c_q$ given by:
\begin{align}
 \xi_q^c &\sim \frac{d_e}{N_e} \sim ( p_b - p_b^c)^{d_f \nu} \quad \textrm{as }p_b \downarrow p_b^{c},  \label{Eq:xicrit}
\end{align}
where we have assumed that $d_e \sim O(1)$, as there is a finite density of points on the lattice where the puddles nearly connect. 
The generic phase boundary has the shape shown in Fig.~\ref{Fig:2dphasediagram}.

There are a few possibilities for the nature of the phase boundary at non-zero $\xi_q$.
The boundary could be first order.
Assuming that it is continuous, however, the line of critical points should lie in the generic Floquet universality class with no symmetries.
By analogy with the semi-classical limit of the integer quantum Hall transition \cite{prange1987quantum,Trugman:1983aa}, we do not expect this to coincide with classical percolation, as at $\xi_q=0$.
Indeed, quantum tunneling across saddle points is well-known to increase the correlation length exponent of the Hall transition \cite{Chalker:1988aa}. 
We therefore conjecture that the same inequality holds for the correlation length exponents of the MBL transition.
We leave the construction of a detailed generic critical theory for future work.

It is clear that the delocalized non-ergodic phase remains delocalized under generic perturbations. However, it is likely to become fully ergodic. 
Any local IOM interacts with the infinite percolating ergodic cluster with some finite strength for any $\xi_q > 0$. However, the level spacing on the infinite cluster vanishes.
Thus, by the arguments above, every local IOM hybridizes with the percolating cluster and delocalizes.

Finally, we provide evidence in $d=1$ that 
(i) localization persists away from the Clifford limit, and 
(ii) there is a transition at $\xi_q > 0$. 
Numerically exact time evolution is limited to small system sizes $N=6-12$; we therefore work deep in the localized regime at $p_b=1$.
A layer of random single qubit rotations provides the perturbation to the Clifford brick-layer circuit. 
The rotation on qubit $i$ is given by $V_i = e^{-i \epsilon_i \vec{\sigma}\cdot\hat{n}/2}$, where $\vec{\sigma}$ is the vector of Pauli operators, $\hat{n}$ is a unit vector on the Bloch sphere with polar angle $\theta$ and azimuthal angle $\pi/4$, and 
\begin{align}
\epsilon_i \in \pm \frac{\pi}{4} + \left [ \frac{-\pi}{12} , \frac{\pi}{12}\right].
\end{align}
is uniformly chosen.

This particular model for quenched disorder can be motivated as follows.
Recall that $[Z_i, U]=0$ for all $i$ at $p_b=1$ in the Clifford limit.
The single-qubit rotations dress the IOM $Z_i$.
If the axis of rotation is slightly tilted from the $z$ axis by an angle $\theta$, we expect the dressing to be quasi-local; this leads to the generic MBL phase.
At larger $\theta>\theta_c$, the dressing is non-local and the circuit has no local IOM. 
This should be the delocalized ergodic phase.

As before, we perform global quenches from the anti-ferromagnetic state in the $z$-basis and study the disorder-averaged saturated entanglement entropy $[S_A(\infty)]$ between the two halves of the chain after $t \approx 2^{24}$ time steps. 
We find that $[S_A(\infty)]$ is much smaller than $N_A=N/2$ at small $\theta$ while it approaches $N/2 - c$ near $\theta = \pi/2$ (not plotted). 
This latter scaling is consistent with a fully ergodic volume-law; the constant correction $c \approx \log_2 e$ as proven in \cite{Page:1993zf}.
However, $[S_A(\infty)]$ increases with $N$ in both phases and the final state entropy density is not a sensitive enough measure at these system sizes to differentiate them.

The standard deviation of $S_A(\infty)$,
\begin{align}
\Delta S \equiv \sqrt{[( \delta S_A(\infty))^2]},
\end{align}
however, behaves quite differently in the localized and delocalized phases \cite{Kjall:2014aa}. 
In the MBL phase, the final entanglement entropy after the quench comes from the dephasing of each of the $N_A$ local IOMs within region $A$. 
As each contributes an $O(1)$ number to $S_A(\infty)$, central limiting behavior implies that $\Delta S \sim \sqrt{N}$.
In the delocalized phase, however, eigenstate thermalization suggests that there should be only exponentially small fluctuations of $S_A$ with system size \cite{Deutsch:1991ss,Srednicki:1994dw}. 
Figure.~\ref{Fig:Fig4} shows the finite-size flow of this quantity as a function of $\theta$, and is clearly consistent with a transition at $\theta_c \approx 0.3 \pi$, although the limited sizes prohibit systematic extraction of the $N$ dependence of the flow.


We note that a localization length $\xi_q$ can be extracted from the residual polarization in the same experiment (not plotted), but its analysis is plagued by the same finite-size effects as the study in Ref.~\cite{Chandran:2014db}.
At the smallest value of $\theta/\pi=0.05$, the finite size flow is not visible within error bars because $[S_A(\infty)]$ is itself very small. 

\emph{Conclusion---}
Periodic quantum circuits provide a useful perspective on localization. 
They distill the underlying entanglement transition from the complicated epiphenomena associated with globally conserved currents \cite{Huse:2013aa,Pekker:2014aa,Vosk:2014aa,Kjall:2014aa,Chandran:2014aa,Bahri:2013qr,Nandkishore:2014aa}.
Here, we identify disordered Clifford circuits as a tractable semi-classical limit with a delocalization transition that maps precisely to percolation, a well-studied beast \cite{Sokolowska:2004aa}.
The percolation exponents show up in dynamical properties such as the growth and saturation of entanglement entropy after a global quench.
We conjecture they also lower bound the correlation length exponent in the generic case with quantum tunneling, just as in the integer quantum Hall plateau transition.
To prove such a result, it would be very interesting to develop a network model for the interacting transition, perhaps in analogy with the recent work in one-dimension \cite{Vosk:2014eu}.
It would also be quite interesting to reintroduce global conservation laws, as these would likely lead to distinct universality classes for the delocalization transition.

\emph{Acknowledgments---} The authors would like to thank D. Abanin, S.L. Sondhi and especially D. Gottesman for helpful discussions and comments. Research at Perimeter Institute is supported by the Government of Canada through Industry Canada and by the Province of Ontario through the Ministry of Research and Innovation. 

\bibliography{master}

\end{document}